\documentclass[aps,prl,showpacs,twocolumn,amsmath,amssymb]{revtex4}
\usepackage{graphicx}
\usepackage{amsmath}
\usepackage{amsfonts}

\renewcommand{\iota}{Groupe d'Optique Atomique, Laboratoire Charles Fabry
de l'Institut d'Optique,\\ UMR 8501 du CNRS,\\ B\^{a}t. 503,
Campus universitaire d'Orsay,\\ 91403 ORSAY CEDEX,
FRANCE}

\begin{document}

\title{Beam quality of a non-ideal atom laser}

\author{J.-F. Riou}
\email{Jean-Felix.Riou@iota.u-psud.fr}
\homepage{http://atomoptic.iota.u-psud.fr}
\affiliation{\iota}
\author{W. Guerin}
\affiliation{\iota}
\author{Y. Le Coq\footnote{Present address:~NIST, Mailcode 847.10, 325 Broadway, Boulder, CO 80305-3328 (U.S.A.)}}
\affiliation{\iota}
\author{M. Fauquembergue}
\affiliation{\iota}
\author{V. Josse}
\affiliation{\iota}
\author{P. Bouyer}
\affiliation{\iota}
\author{A. Aspect}
\affiliation{\iota}

\date{\today}

\begin{abstract}
We study the propagation of a non-interacting atom laser distorted by the strong lensing effect of the Bose-Einstein Condensate (BEC) from which it is outcoupled. We observe a transverse structure containing
caustics that vary with the density within the residing BEC. Using WKB approximation, Fresnel-Kirchhoff integral formalism and ABCD matrices, we are able to describe analytically the atom laser propagation. This allows us to
characterize the quality of the non-ideal atom laser beam by a generalized M$^2$ factor defined in
analogy to photon lasers. Finally we measure this quality factor for different lensing effects.
\end{abstract}


\maketitle

Optical lasers have had an enormous impact on science and
technology, due to their high brightness and coherence. The high
spatial quality of the beam and the little spread when propagating in
the far-field enable applications ranging from the focusing onto
tiny spots and optical lithography \cite{ito:2000} to collimation
over astronomic distances \cite{Bender:1973}. In atomic physics,
Bose-Einstein condensates (BEC) of trapped atoms \cite{BEC:1995} are
an atomic equivalent to photons stored in a single mode of an
optical cavity, from which a coherent matter wave (atom laser) can be extracted
\cite{Mewes:1997, Bloch:1999}. The possibility of creating
continuous atom laser \cite{Chikkatur:2002} promises
spectacular improvements in future applications \cite{Bouyer:1997,Wang:2005,Shin:2005,Sligte:2004,LeCoq:2005} where perfect collimation or strong focusing \cite{Bloch:2001,Schvarchuck:2002,Arnold:2004} are of prior importance. Nevertheless these properties
depend drastically on whether the diffraction limit can be achieved. Thus, characterizing the deviation from this limit is, as for optical lasers \cite{Siegman:1993},
of crucial importance. For example, thermal lensing effects in optical laser cavities, which cause significant decollimation, can also induce aberrations that degrade the transverse profile. In atom optics, a trapped BEC
weakly interacting with the outcoupled atom-laser beam acts as an effective
thin-lens which leads to the divergence of the atom laser
\cite{LeCoq:2001} without affecting the diffraction limit. When the lensing effect increases, dramatic
degradations of the beam are predicted \cite{Busch:2002}, with the apparition of caustics on the edge of the beam.
\newline

\begin{figure}[htb]
\centering \scalebox{.75}{\includegraphics{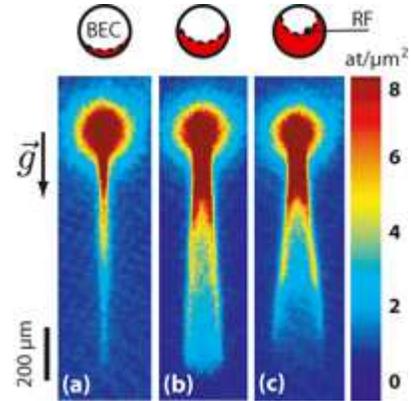}}
\caption{Absorption images of a non-ideal atom laser, corresponding
to density integration along the elongated axis of the BEC. The
figures correspond to different height of RF-outcoupler detunings with respect
to the bottom of the BEC:~(a) $0.37~\mu$m (b) $2.22~\mu$m (c) $3.55~\mu$m. The graph
above shows the RF-outcoupler (dashed line) and the BEC slice
(red) which is crossed by the atom laser. This results in the
observation of caustics. The field of view is $\mathrm{350\, \mu
m\times 1200\, \mu m}$ for each image.} \label{fig1}
\end{figure}

In order to quantitatively qualify the \textit{atom-laser beam quality},
it is tempting to take advantage of the methods developed in optics
to deal with non-ideal laser beams \textit{i.e.} above the diffraction limit. Following the initial work of
Siegman \cite{Siegman:1991} who introduced the quality factor M$^2$
which is proportional to the space-beam-width (divergence $\times$
size) product at the waist, it is natural to extend its definition
to atom optics as
\begin{equation}
\Delta x \;\Delta k_x\; =\; \frac{\mathrm{M^2}}{2},
\label{heisenberg}
\end{equation}
where $\Delta x$ and $\Delta k_x=\Delta
p_x/\hbar$ characterize respectively the size and the divergence along $x$ ($\Delta p_x$ is the width of the momentum distribution). 
Equation (\ref{heisenberg}) plays the same role as the
Heisenberg dispersion relation:~it expresses how many times the
beam deviates from the diffraction limit.

In this letter, we experimentally and theoretically study the
quality factor M$^{2}$ of a non-ideal, non-interacting atom-laser beam. First, we
present our experimental investigation of the structures that appear
in the transverse profile \cite{Kohl:2005}. We
show that they are induced by the strong lensing effect due to the interactions between the trapped BEC and the outcoupled beam. Then, using an
approach based on the WKB approximation and the Fresnel-Kirchhoff integral formalism, we
are able to calculate analytical profiles which agree with our
experimental observations. This allows us to generalize
concepts introduced in \cite{Siegman:1991} for photon laser 
and to calculate the quality factor M$^2$. This parameter can then be used in combination with the paraxial ABCD matrices \cite{Belanger:1991} to describe the propagation of the non-ideal beam via the evolution of the rms width.
Finally, we present a study of the M$^2$ quality factor as a
function of the thickness of the BEC-induced output lens.

Our experiment produces atom lasers obtained by
radio frequency (RF) outcoupling from a BEC \cite{Bloch:1999,Gerbier:2001}. The
experimental setup for creating condensates of
$\mathrm{{}^{87}Rb}$ is described in detail in
\cite{Fauquembergue}. Briefly, a Zeeman-slowed atomic beam loads a
magneto-optical trap in a glass cell. About $2\times 10^8$ atoms are
transferred in the $\left|F,m_{F}\right\rangle=\left|1,-1\right\rangle$ state to a
Ioffe-Pritchard magnetic trap , which is subsequently compressed to
oscillation frequencies of $\omega_{y }=2 \pi \times 8$ Hz and
$\omega_{x,z}=2\pi\times 330$ Hz in the dipole and quadrupole
directions respectively. A 25 s  RF-induced
evaporative cooling ramp results in a pure condensate of $N=10^6$
atoms, cigar-shaped along the $y$ axis.

\begin{figure}[htb]
\centering \scalebox{.75}{\includegraphics{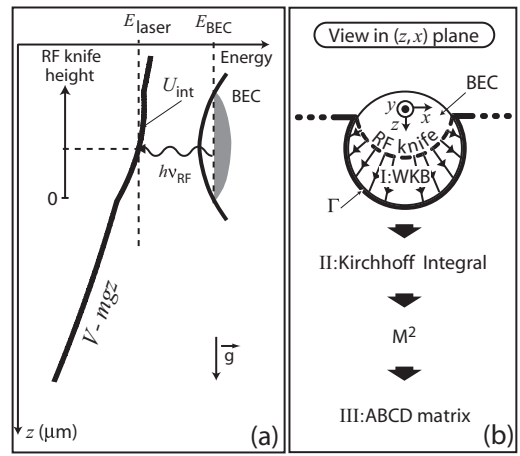}} \caption{(a)
Principle of the RF-outcoupler :~the radio-frequency $\nu_{RF}=
(E_{\mathrm{BEC}}-E_{\mathrm{laser}})/h$ selects the initial
position of the extracted atom-laser beam. The laser is then subjected to the
condensate mean-field potential $U_{\rm int}$, to the quadratic Zeeman 
effect $V$  and to gravity $-mgz$. For the sake of clarity, $V$ has been exaggerated on the graph.
(b) Representation of the two-dimensional
theoretical treatment: (I) inside the condensate, phase integral along atomic paths determines the laser wavefront at the BEC output (WKB approximation). (II) A Fresnel-Kirchhoff integral on the contour $\Gamma$ is used to calculate the
stationary laser wavefunction at any point below the condensate. (III) As soon as
the beam enters the paraxial regime, we calculate the $\mathrm{M^2}$
quality factor and use ABCD matrix
formalism.} \label{fig2}
\end{figure}

The atom laser is extracted from the BEC by applying a RF field a
few kHz above the bottom of the trap, in order to couple the trapped state to
the weakly anti-trapped state $\left|1,0\right\rangle$. The extracted atom laser beam falls under the effect of both gravity $-mgz$
and second order Zeeman effect $V=-m\omega^2(x^2+z^2)/2$ \cite{Desruelles:1999}  with
$\omega = 2\pi \times 20$ Hz (see Fig. \ref{fig2}a). The
RF-outcoupler amplitude is weak enough to avoid perturbation of the
condensate so that the laser dynamics is quasi-stationary
\cite{Gerbier:2001} and the resulting atom flux is low enough to avoid interactions within the propagating beam. Since the BEC is displaced vertically by the
gravitational sag, the value of the RF-outcoupler frequency
$\nu_{\rm RF}$ defines the height where the laser is extracted
\cite{Bloch:1999}. After 10 ms of operation, the
fields are switched off and absorption imaging is taken after 1 ms of
free fall with a measured spatial resolution of 6 $\mu$m. The line
of sight is along the weak $y$ axis so that we observe the
transverse profile of the atom laser in the ($z,x$) plane.

Typical images are shown in figure \ref{fig1}. Transverse
structures, similar to the predictions in \cite{Busch:2002}, are clearly visible in figures \ref{fig1}b and \ref{fig1}c. The
laser beam quality degrades as the RF-outcoupler is higher in the BEC (i.e. the
laser beam crosses more condensate), supporting the interpretation that this effect is due to the
strong repulsive interaction between the BEC and the laser. 
This effect can be understood with a semi-classical picture. 
The mean-field interaction results in an inverted harmonic potential
of frequencies $\omega_{i}$ (in the directions $i=x,y,z$) which, in the Thomas-Fermi regime, are
fixed by the magnetic confinement \cite{Stringari,LeCoq:2001}. 
The interaction potential expels the atoms transversally, as illustrated in
Fig. \ref{fig2}b. Because of the finite size of the condensate, the
trajectories initially at the center of the beam experience more
mean-field repulsion than the ones initially at the border.
This results in accumulation of trajectories at the edge of the atom laser beam \cite{Busch:2002}, in a similar manner to caustics in optics. This
picture enables a clear physical understanding of the behaviour
observed in figure \ref{fig1}:~if $\nu_{\rm RF}$ is chosen so that
extraction is located at the bottom of the BEC (Fig. \ref{fig1}a),
the lensing effect is negligible and one gets a collimated
beam. As the RF outcoupler moves upwards
(Fig. \ref{fig1}b and \ref{fig1}c), a thicker part of the condensate acts on the laser and defocusing, then caustics  appear.  We verified that when sufficiently decreasing the
transverse confinement of the trapped BEC, {\it i.e.} making the interaction with the outcoupled atoms negligible, the atom laser is collimated at any RF
value \cite{Nvar}.

In order to describe quantitatively the details of the profiles
of the non-ideal atom laser one could solve numerically the Gross-Pitaevskii equation (GPE) \cite{Busch:2002,Kohl:2005}. As we show here, another approach is possible, using approximations initially developed in the context of photon optics and extended to atom
optics. These approximations allow calculation of the atom-laser
propagation together with the characterization of its rms width evolution by means of the
quality factor M$^2$, in combination with ABCD matrices. Note however that these matrices can only be used in the paraxial regime, {\it i.e.} when the transverse
kinetic energy is smaller than the longitudinal one \cite{LeCoq:2001}. This condition does not hold in the vicinity of the BEC, and thus we split the atom laser evolution into three steps (Fig. \ref{fig2}b) where we use different formalisms in close analogy with optics: (I) WKB inside the condensate (eikonal), (II) Fresnel-Kirchhoff integral outside the condensate and, (III) paraxial ABCD matrices after sufficient height of fall.
\begin{figure}[htb]
\centering \scalebox{.8}{\includegraphics{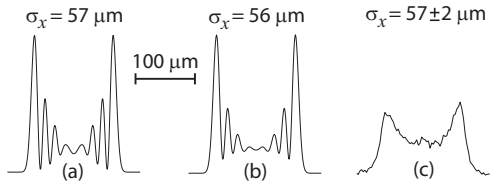}}
\caption{Beam profile after 600 $\mu$m of propagation for a outcoupler height of 3.55 $\mu$m: (a) calculated in the central plane $y=0$; (b) resulting from the integration along the line of sight $y$; (c) obtained experimentally. Deviations between the curves (b) and (c) can be attributed to imperfections in the imaging process and bias fluctuations. Note that for all three profiles, the overall shape is preserved and the calculated rms width in the central plane is in agreement with the measured one within experimental errors.} \label{fig3}
\end{figure}
In all the following, we do not include interactions within the atom laser since they are negligible for the very dilute beam considered in this letter. In addition, the condensate is elongated
along the $y$ axis, so that the forces along this direction are negligible
and we consider only the dynamics in the ($z,x$) plane.

First, we consider that the atoms extracted from the condensate by the
RF-outcoupler start at zero
velocity from $\mathbf{ r}_0$. In region (I) the total potential (resulting from gravity and interactions with the trapped BEC) is cylindrically symmetric along the $y$ axis.
The trajectories are thus straight lines (see Fig. \ref{fig2}b). The beam profile $\psi(\mathbf{ r}_1)$ at the BEC border is obtained thanks to the WKB approximation, by integrating the phase along classical paths of duration $\tau_0$ 
\begin{equation}
\psi(\mathbf{ r}_1)\propto \frac{1}{\sqrt{\sinh{(2\omega_{z}\tau_0})}} e^{i
\int_{\mathbf{ r}_0}^{\mathbf{ r}_1} \mathbf{ k}(\mathbf{ r})\cdot
d\mathbf{ r}}\psi_{\mathrm{BEC}}(\mathbf{ r}_0), \label{BECfrontier}
\end{equation}
where $\psi_{\mathrm{BEC}}$ is the condensate wavefunction and the prefactor ensures the conservation of the flux.

The atom laser being in a quasi-stationary state, the wavefunction satisfies in region (II) the time-independent Schr\"{o}dinger equation at a given energy $E$ in the
potential $V-mgz$. Since this equation is analogous to the Helmoltz
equation \cite{BornWolf, Henkel} in optics, the Fresnel-Kirchhoff integral
formalism can be generalized to atom optics as \cite{Borde}
\begin{equation}
\psi(\mathbf{ r})\propto \oint_{\Gamma} d\mathbf{ l}_1 \cdot \left[G_E
\mathbf{ \nabla}_1\psi(\mathbf{ r}_1)-\psi(\mathbf{ r}_1)
\mathbf{ \nabla}_1G_E\right].\label{Kirchoff}
\end{equation}
We take the contour $\Gamma$ along the condensate border and close it at infinity (see Fig. \ref{fig2}b ).
The time-independent
Green's function $G_E(\mathbf{ r},\mathbf{ r}_1)$ is analytically
evaluated from the time-domain Fourier
 transform of the Feynman propagator $K(\mathbf{ r},\mathbf{ r}_1,\tau)$ \cite{Borde,Feynman} calculated
 by means of the van Vleck formula \cite{VanVleck}.
 Using $\psi(\mathbf{ r_1})$ from equation (\ref{BECfrontier}),
 the wavefunction $\psi(\mathbf{ r})$ is then known at any location $\mathbf{ r}$. 
We verified the excellent agreement of this model with a numerical integration of the GPE including intra-laser interactions, thus confirming that they remain negligible throughout propagation.

The method using a Kirchhoff integral is demanded only for the early stages of the propagation. As soon as the paraxial
approximation becomes valid, the ABCD matrix formalism can be used to describe the propagation of the beam. An example of a profile calculated in the central plane ($y=0$) is presented in figure \ref{fig3}a. In figure \ref{fig3}b 
we add the profiles of all $y$ planes taking into account the measured resolution of the imaging system. In figure \ref{fig3}c we present the corresponding experimental profile. The overall shape is in good agreement with theory, and the  differences can be explained by imperfections in the imaging process and bias fluctuations during outcoupling. The rms size of the three profiles, which depend very smoothly on the details of the structure, are the same within experimental uncertainties. Thus, hereafter, as in the case of propagation of optical laser using the M$^2$ parameter, we only consider the evolution of the rms size of the atom laser. 

To calculate the beam width change in the paraxial regime, we define, following \cite{Siegman:1991}, a generalized complex radius of curvature
\begin{equation}
\frac{1}{q(\xi)}={\cal C}(\xi)+\frac{i\mathrm{M^2}}{2\sigma_x^2(\xi)},
\label{q}
\end{equation}
where $\sigma_x$ is the rms width of the density profile, $\xi(t)$ a reduced variable which
describes the time evolution of the beam such that $\xi=0$ corresponds to the position of the waist. 
Equation (\ref{q}) involves an invariant coefficient, the beam-quality factor M$^{2}$, as defined in Eq. (\ref{heisenberg}).
This coefficient, as well as the effective curvature ${\cal C}(\xi)$, can be extracted from
the wavefront in the paraxial domain, as explained in \cite{Siegmanbook}.
In optics, the generalized complex radius obeys the same ABCD
propagation rules as does a Gaussian beam of the same real beam
size, if the wavelength $\lambda$ is changed to M$^2\lambda$
\cite{Belanger:1991}.  Similarly, the complex radius $q(\xi)$ follows here the
ABCD law for matter-waves \cite{LeCoq:2001}, and we obtain the rms width
\begin{equation}
\sigma_x^2(\xi)=\sigma_{x0}^2\cosh^2(\xi)+\left(\frac{\mathrm{M^2}\hbar}{2m\omega}\right)^2
\frac{\sinh^2(\xi)}{\sigma_{x0}^2}, \label{sigma}
\end{equation}
where $\sigma_{x0}$ is taken at the waist.

\begin{figure}[htb]
\centering \scalebox{.9}{\includegraphics{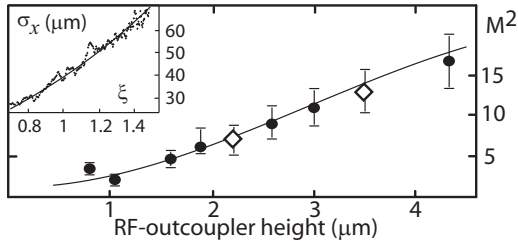}}
\caption{$\mathrm{M^2}$ quality factor vs RF-outcoupler distance from the bottom of the BEC:~theory (solid line), 
experimental points
(circles). The two diamonds represent the $\mathrm{M^2}$  for the two non-ideal
atom lasers shown in figure \ref{fig1}b and \ref{fig1}c. The RF-outcoupler position is
calibrated by the number of outcoupled atoms. Inset:~typical fit
of the laser rms size with the generalized Rayleigh formula
(Eq. \ref{sigma}) for RF-outcoupler position $3.55~\mu$m.} \label{fig4}
\end{figure}

This generalized Rayleigh formula allows us to measure $\mathrm{M^2}$. In the inset of figure \ref{fig4}, 
the evolution of the transverse rms width $\sigma_x$ versus propagation $\xi$, taken from experimental images, is
compared to the one given by Eq. (\ref{sigma}), where  $\sigma_{x0}$ is calculated with our model.
For a chosen RF-outcoupler position, we fit the variation of the width with a single free parameter M$^2$. We then plot the measured value of M$^2$ vs the ouput coupler height (Fig. \ref{fig4}), and we find good agreement with theory.

In conclusion, we have characterized the transverse profile of an
atom laser. We demonstrated that, in our case, lensing effect when crossing the condensate
is a critical contributor to the observed degradation of the beam. We showed that
the beam-quality factor M$^2$, initially introduced by Siegman \cite{Siegman:1991}
for photon laser, is a fruitful concept for describing the propagation of an atom laser beam with ABCD matrices, as well as for characterizing how far an atom laser deviates from the diffraction limit.
For instance, it determines the minimal focusing size that can be achieved 
with atomic lenses provided that interactions in the laser remain negligible \cite{Arnold:2004, comment}. This is of essential importance in view of future applications of coherent matter-waves as, for example, when coupling atom lasers onto guiding structures of atomic chips \cite{EPJD}. In addition, if interactions within atom laser become non negligible, a further treatement could be developed in analogy with the work of \cite{Pare:1992} for non-linear optics.

\acknowledgements
The authors would like to thank S. Rangwala, A. Villing and F. Moron for their help on the experiment, L. Sanchez-Palencia and I.
Bouchoule for fruitful discussions and R. Nyman for careful reading of the manuscript. This work is
supported by CNES (DA:10030054), DGA (contract 9934050 and 0434042), 
LNE, EU (grants
IST-2001-38863, MRTN-CT-2003-505032 and FINAQS STREP), INTAS
(contract 211-855) and ESF (BEC2000+).


\begin{thebibliography}{llml}
\bibitem{Nvar}{We also verified that the lensing effect varies very little with  condensate atom number since the Thomas Fermi radius, 
and thus the part acting on the atom laser varies as $N^{1/5}$ \cite{Stringari}.}
\bibitem{comment} For a focusing size $R$ the transverse kinetic energy deduced from equation (\ref{heisenberg}) is $\left(\mathrm{M}^2\right)^2\hbar^2/4mR^2$ whereas the interaction energy reads $2\pi\hbar^2a {\cal{F}}/v m R^2$, with $a$, $v$ and ${\cal{F}}$ being repectively the scattering length, the on-axis beam velocity and the flux. Interactions are thus negligible if ${\cal{F}}/v\ll \left(\mathrm{M}^2\right)^2/8\pi a$.
\end{thebibliography}
\end{document}